\documentclass[letterpaper,twocolumn,10pt]{article}
\usepackage{usenix}

\usepackage[dvips]{graphicx}
\usepackage{url,multirow}

\usepackage[dvipdfm,CJKbookmarks=true,bookmarksopen=true,colorlinks,citecolor=black,linkcolor=blue,anchorcolor=black,urlcolor=black] {hyperref}
\usepackage{balance}

\begin{document}

%don't want date printed
\date{}

%make title bold and 14 pt font (Latex default is non-bold, 16 pt)

\title{\LARGE \bf Transaction Support over Redis: An Overview}

%for single author (just remove % characters)
\author{
{\rm Yuqing Zhu, Jianxun Liu, Mengying Guo, Wenlong Ma, Yungang Bao}\\%
Institute of Computing Technology, Chinese Academy of Sciences
% copy the following lines to add more authors
% \and
% {\rm Name}\\
%Name Institution
} % end author

\maketitle

% Use the following at camera-ready time to suppress page numbers.
% Comment it out when you first submit the paper for review.
\thispagestyle{empty}

%Your Abstract Text Goes Here.  Just a few facts. Whet our appetites.
This document serves as an overview for describing how to support cross-node transactions over a Redis cluster.

Redis is a key-value in-memory store. It supports the common interface for data access. Besides, it supports the transaction-like interfaces of MGET/MSET and MULTI/EXEC locally on a Redis node. The two transaction-like interfaces enable useful and handy implementations of applications. But unfortunately they are not supported in the cluster mode of Redis.

\begin{figure}[!h]
\vspace{-6pt}
      \centering
      \includegraphics[width=0.45\textwidth]{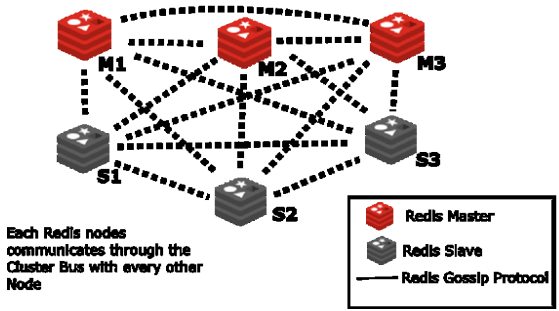}\vspace{-3pt}
      \caption{A six-node Redis cluster.}\vspace{-9pt}
      \label{fig:redisCluster} %% label for entire figure
\end{figure}
The cluster mode of Redis guarantees scalability through data partition and high availability through the master-slave replication. A typical minimum Redis cluster setup is to have a three Master Redis nodes with each master node replicated with a single Redis slave instance node, as illustrated in Figure~\ref{fig:redisCluster}. The keys stored in a Redis cluster are typically partitioned into orthogonal segments, each of which is stored by a Mater node. To support the transaction-like accesses on arbitrary keys, cross-node transactions must be supported over a Redis Cluster.

As high availability is widely demanded by online applications, we therefore propose to implement transactions with ACIA (Atomicity, Consistency, Isolation and Availability) properties~\cite{acia} in the current Redis distribution. To guarantee atomicity, we adopt the HACommit atomic commit protocol~\cite{hacommit}. To guarantee isolation, we propose to implement concurrency control through consensus reduction~\cite{zhuIso}; or, if the read-committed isolation level is allowed, we can simply adopt the write-on-commit mechanism. To guarantee high availability, many replica control methods~\cite{recods,stepReplica} can be employed while the master-slave replication of Redis is also a feasible choice given an optimistic assumption on the deployment. The consistency property is generally specified through legal rules. We leave the support of legal rule specifications and consistency choices for the future version of ACIA transaction systems.

\begin{figure}[!h]
\vspace{-6pt}
      \centering
      \includegraphics[width=0.3\textwidth]{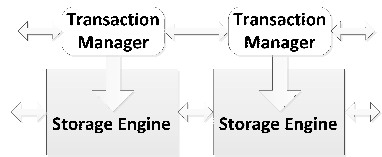}\vspace{-3pt}
      \caption{The architecture for the cross-node transaction implementation.}\vspace{-9pt}
      \label{fig:arch} %% label for entire figure
\end{figure}
\begin{figure}[!h]
\vspace{-6pt}
      \centering
      \includegraphics[width=0.3\textwidth]{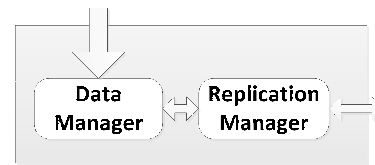}\vspace{-3pt}
      \caption{The typical architecture of a highly-available storage engine.}\vspace{-9pt}
      \label{fig:availability} %% label for entire figure
\end{figure}

With the above implementation choices, the overall system architecture is depicted in Figure~\ref{fig:arch}. The communication protocol between transaction managers are the atomic commit protocol HACommit.  The storage engine is the Redis system. The communications between storage engines follow the pattern implemented by the cluster-mode Redis. In our architecture, the storage engine is one guaranteeing high availability, as demonstrated in Figure~\ref{fig:availability}. Replication is an inevitable component in the highly-available storage engine, while given the master-slave replication of Redis, HACommit is the fastest atomic commit protocol to choose.

\begin{figure}[!h]
\vspace{-6pt}
      \centering
      \includegraphics[width=0.3\textwidth]{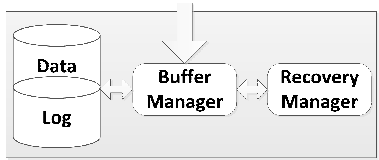}\vspace{-3pt}
      \caption{The typical architecture of a durable storage engine.}\vspace{-9pt}
      \label{fig:durability} %% label for entire figure
\end{figure}

Note that the highly-available storage engine can also be replaced by the durable storage engine in Figure~\ref{fig:durability} if needed. In implementations, we need only to remove the communications with replicas and assume the write-ahead logging in the critical path of writes involved in the transaction processing.

Now, we have described how transactions can be supported over a storage engine like Redis. Given that transactions are implemented over Redis, the transaction-like interfaces of MGET/MSET and MULTI/EXEC can both be supported. Further information can be found on the Web Page\footnote{https://sites.google.com/site/zhuyuqing/acia}.

\textbf{Acknowledgments} This work is supported in part by the State Key Development Program for Basic Research of China (Grant No. 2014CB340402) and the National Natural Science Foundation of China (Grant No. 61303054).

\balance

{\footnotesize \bibliographystyle{acm}
\bibliography{ref}}

%\theendnotes

\end{document}